\documentclass[letterpaper]{article}
\usepackage{times}
\usepackage{aaai}
\usepackage{helvet}
\usepackage{courier}
\usepackage{fancyhdr}
\fancypagestyle{plain}{\pagestyle{fancy}}
\usepackage[numbers,sort&compress]{natbib}
\usepackage{caption}
\usepackage{xurl}
\usepackage[breaklinks=true]{hyperref}
\usepackage{enumitem}
\usepackage{needspace}
\usepackage{graphicx}  % For \resizebox
\usepackage{booktabs}  % For \toprule, \midrule, and \bottomrule
\usepackage{array}
\usepackage{xcolor}
\usepackage{textcomp}  % Additional text symbols
\usepackage{amssymb}   % Mathematical symbols
\begin{document}
\title{Acceptable Use Policies for Foundation Models}
\author{Kevin Klyman\\ Stanford University, Center for Research on Foundation Models \\ Harvard University, Belfer Center for Science and International Affairs \\ kklyman@stanford.edu}
\maketitle
\begin{abstract}
\begin{quote}
As foundation models have accumulated hundreds of millions of users, developers have begun to take steps to prevent harmful types of uses. 
%organizations that develop such models
One salient intervention that foundation model developers adopt is \textit{acceptable use policies}---legally binding policies that prohibit users from using a model for specific purposes. 
This paper identifies acceptable use policies from 30 foundation model developers, analyzes the use restrictions they contain, and argues that acceptable use policies are an important lens for understanding the regulation of foundation models.  
Taken together, developers' acceptable use policies include 127 distinct use restrictions; the wide variety in the number and type of use restrictions may create fragmentation across the AI supply chain.

Developers also employ acceptable use policies to prevent competitors or specific industries from making use of their models.
Developers alone decide what constitutes acceptable use, and rarely provide transparency about how they enforce their policies. 
In practice, acceptable use policies are difficult to enforce, and scrupulous enforcement can act as a barrier to researcher access and limit beneficial uses of foundation models. 
Nevertheless, acceptable use policies for foundation models are an early example of self-regulation that have a significant impact on the market for foundation models and the overall AI ecosystem. %claim to be the first?

%Nevertheless, they have emerged as one of the most common mechanisms for limiting the harm of foundation models.
\end{quote}
\end{abstract}
\vspace{-3pt}
\section{1. Introduction}
\noindent Policymakers hoping to regulate foundation models have focused on preventing specific objectionable uses of AI systems, such as the creation of bioweapons \citep{RR-A2977-2}, deepfakes \citep{nist2024reducing}, and child sexual abuse material \citep{thorn2024csam}. 
Effectively blocking these uses can be difficult in the case of foundation models---large AI models trained on broad data that can be adapted to a wide range of downstream tasks---as they are general-purpose technologies that in principle can be used to generate any type of content \citep{bommasani2022opportunities}. 
Yet developers of foundation models have been proactive, 
adopting broad policies as part of their terms of service or model licenses that prohibit many potentially dangerous uses of the technology. 

Foundation model developers have taken several approaches to adopting legally binding use restrictions. 
\cite{mcduff2024standardization} find that developers of open-weight foundation models increasingly distribute these models with licenses that include a standardized set of behavioral use restrictions.
Developers of closed-weight models have also restricted how users can make use of their models, often via terms of service agreements that prohibit generating specific categories of content \citep{bommasani2023foundation}. 
Developers often refer to policies that include legally binding use restrictions on foundation models as acceptable use policies (AUPs), as they determine the domains of use that are acceptable and prohibited.

This paper collates and analyzes the acceptable use policies of 30 foundation model developers in order to assess their impact. It addresses the following question: what do acceptable use policies reveal about the ways that foundation model developers seek to regulate end-user behavior, and how do they impact the foundation model ecosystem?

The paper proceeds as follows: 
Section 2 provides background on acceptable use policies for foundation models, comparing them to similar policies for other technologies and to documents like model cards (which list out-of-scope uses but are not legally binding). 
Section 3 describes the methodology used to identify acceptable use policies and analyze their content.
Section 4 analyzes the differences between developers' policies in terms of prohibited content and restrictions on types of end use. %\textbf{finding?????}
Section 5 outlines difficulties in policy enforcement and potential downsides from strict enforcement. 
Section 6 discusses developers' decision-making power, how gaps in use restrictions may facilitate misuse, and how acceptable use policies shape the foundation model market.
Section 7 identifies areas for future work. 

\section{2. Background}
\textbf{2.1 What is an acceptable use policy?}

\noindent Acceptable use policies are common across digital technologies \citep{doi:https://doi.org/10.1002/9781118978238.ieml0001}. 
Providers of public access computers \citep{robinson2020to}, websites \citep{steward2000internet, WeidmanGrossklags2019}, and digital platforms \citep{10.1145/502269.502302, pater2016characterizations} have long adopted acceptable use policies that articulate how their terms of service restrict what users can and cannot do with their products and services. 
While enforcement of these policies is uneven, restrictions on specific uses of digital technologies are widespread \citep{doherty2011reinforcing, RUIGHAVER2010731}.

The acceptable use policies of social media companies \citep{election2021longfuse}, cloud service providers \citep{https://doi.org/10.1002/itl2.84}, and content delivery networks \citep{https://doi.org/10.1002/poi3.370} have received scrutiny as they constrain the behavior of hundreds of millions of users. %\textbf{INSERT CONTENT POLICY ANALYSIS}
Acceptable use policies adopted by employers, which limit employees' use of company-provided technologies \citep{LI2010635}, schools, which limit students' use of the internet \citep{https://doi.org/10.1002/meet.2011.14504801044}, and public libraries, which limit the public's use of public access computers \citep{mcmenemy2019public}, have come into focus as issues related to enforcement arise.   

In the context of foundation model development, an acceptable use policy is a policy from a developer that determines how a foundation model can or cannot be used. 
Acceptable use policies restrict the use of foundation models by detailing the types of content users are prohibited from generating as well as domains of prohibited use.\footnote{This differs from the case of non-generative technologies, where restrictions focus on a user's input not a system's output.}
Developers make these restrictions legally binding by including acceptable use policies in terms of service agreements or in copyright licenses for their foundation models.

Acceptable use policies typically aim to prevent users from using a foundation model to generate content that may violate the law or otherwise cause harm.\footnote{Acceptable use policies for 3D printers, another generative, general-purpose technology with the potential to cause real-world harm \citep{Beyer2014}, are among the best analogue for the case of foundation models. 
Public libraries have adopted AUPs that prohibit 3D printing of ghost guns, sex toys, and swastikas, for instance \citep{https://doi.org/10.1002/bul2.2015.1720420113, Minow2016}.} 
They accomplish this by listing specific subcategories of violative content and authorizing model developers to punish users who generate such content by, for example, limiting the number of queries users can issue or banning a user’s account.

Acceptable use policies relate to how foundation models are built in important ways. 
For example, developers frequently filter training data to remove content relevant to requests that would violate their acceptable use policies. 
OpenAI’s GPT-4 technical report states: “We reduced the prevalence of certain kinds of content that violate our usage policies (such as inappropriate erotic content) in our pre-training dataset, and fine-tuned the model to refuse certain instructions such as direct requests for illicit advice” \citep{openai2024gpt4}.

In addition, many developers state that the purpose of reinforcement learning from human feedback (RLHF) is to make their foundation models less likely to generate outputs that would violate their acceptable use policies \citep{lambert2023history}. 
Meta’s technical report for Llama 2 notes that the risks RLHF was intended to mitigate include “illicit and criminal activities (e.g., terrorism, theft, human trafficking); hateful and harmful activities (e.g., defamation, self-harm, eating disorders, discrimination); and unqualified advice (e.g., medical advice, financial advice, legal advice),” which correspond to the acceptable use policy in Llama 2’s license \citep{touvron2023llama}. 
Anthropic’s model card for Claude 3 similarly says “We developed refusals evaluations to help test the helpfulness aspect of Claude models, measuring where the model unhelpfully refuses to answer a harmless prompt, i.e. where it incorrectly categorizes a prompt as unsafe (violating our AUP) and therefore refuses to answer” \citep{anthropic2024claude3}.
\newline

\noindent \textbf{2.2 How do AUPs differ from other similar documents?}

\noindent Acceptable use policies are not the only way developers restrict use of their models. 
Other policy-related mechanisms that developers implement to restrict model use include: 

\begin{itemize}[itemsep=0pt,parsep=0pt,topsep=0pt]
    \item \textit{Model Cards}: Model cards, which are published alongside machine learning models, provide essential information about models such as their intended uses and out-of-scope uses \citep{10.1145/3287560.3287596}. However, model cards are not enforceable contracts, and they are not generally referenced in model licenses or developers' terms of service; as a result, out-of-scope uses do not rise to the same level as prohibited uses in an acceptable use policy \citep{liang2024whatsdocumentedaisystematic}. %Cite more model card analyses
    \item \textit{Model Behavior Policies}: Model behavior policies determine what a model can or cannot do   \citep{openai2024modelspec, google2024geminispec, bai2022constitutionalaiharmlessnessai}. %ADD GOOGLE'S BEHAVIOR POLICY 
    While acceptable use policies apply to user behavior, model behavior policies apply to the behavior of the model itself \citep{bommasani2023foundation}. A model behavior policy is one way of embedding an acceptable use policy into a model; methods for imposing a model behavior policy include using RLHF to cause the model to be more likely to refuse violative prompts or employing a safety classifier at inference time to filter violative model outputs \citep{christiano2023deepreinforcementlearninghuman, brahman2024artsayingnocontextual, inan2023llamaguardllmbasedinputoutput}. Model behavior policies are generally broader than acceptable use policies; for instance, many developers fine-tune their models to produce more polite responses, though they do not block users from generating impolite responses \citep{schneider2024exploringhumanllmconversationsmental, Priya_2024}.
    \item \textit{Third party contracts}: Foundation model developers frequently partner with other firms to disseminate foundation models \citep{cen2023ai}. 
    These include cloud service providers (e.g., AWS, Azure, GCP), platform providers (e.g., Scale AI, Nvidia), database providers (e.g., Salesforce, Oracle), and model distributors (e.g., Together, Quora) \citep{srikumar2024risk}. 
    Custom contracts with third party providers of a developer’s foundation models often include use restrictions, but the extent to which companies’ acceptable use policies are altered via these partnership agreements is unclear. \\ 
\end{itemize}
\noindent \textbf{2.3 Norms and laws on acceptable use policies}

\noindent Although generative AI is a nascent industry, norms have begun to emerge around use restrictions for foundation models. 
\cite{cohere2022best} wrote in their “Best practices for deploying language models” that organizations should “[p]ublish usage guidelines and terms of use of LLMs in a way that prohibits material harm...such as through spam, fraud, or astroturfing.” 
Developers of open-weight foundation models often adopt the same acceptable use policies by reusing the same model licenses. For example, more than 3,000 models on Hugging Face use Meta's Llama 2 license \citep{mcduff2024standardization}.

Governments have taken an interest in acceptable use policies, which are a salient effort by foundation model developers to self-regulate \citep{ferretti2022institutionalist}.
Annexes IXa and IXb of the EU AI Act require that all providers of general-purpose AI models disclose the “acceptable use policies [that are] applicable” to both the EU’s AI Office and other firms that integrate the general-purpose AI model into their own AI systems \citep{EU_AI_Act_2024, hacker2023airegulationeuropeai}. 
China’s Interim Measures for the Management of Generative AI Services, which were adopted in July 2023, go a step further by requiring that providers of generative AI services act to prevent users from “using generative AI services to engage in illegal activities…including [by issuing] warnings, limiting functions, and suspending or concluding the provision of services” \citep{China_AI_Interim_Measures_2023, zhang2024promise}. And the US Voluntary AI Commitments require firms to publicly report “domains of appropriate and inappropriate use” as well as any limitations of the model that affect these domains \citep{wh2023voluntary}. 

\begin{table*}[ht!]
\resizebox{\textwidth}{!}{%
\begin{tabular}{>{\raggedright\arraybackslash}p{2cm}>{\raggedright\arraybackslash}p{8cm}lp{1.6cm}p{4cm}llll}
\toprule

\textbf{Developer} & \textbf{Title of Acceptable Use Policy, \newline Section Including Use Restrictions} & \textbf{\parbox[t]{2.5cm}{Model Specific \newline Y/N (Model)}} & \textbf{Policy Document} & \textbf{Flagship Model Series \newline (Output Modality)} & \textbf{HQ} & \textbf{Openness} & \textbf{Ref} \\
\midrule
01.ai & Yi Series Models Community License Agreement v2.1, §2 License and License Restrictions & Y (Yi) & License & Yi (Text) & PRC & Open & [\href{https://huggingface.co/01-ai/Yi-34B/blob/main/LICENSE}{1}] \\
Adept & Terms of Use, §1.1(d) Usage Restrictions & N & TOS & Fuyu (Multimodal) & USA & Open & [\href{https://www.adept.ai/policies/terms-of-use}{2}] \\
Adobe & Generative AI User Guidelines & N & Standalone & Firefly (Image) & USA & Closed & [\href{https://www.adobe.com/legal/licenses-terms/adobe-gen-ai-user-guidelines.html}{3}] \\
AI21 & Usage Guidelines & N & Standalone & Jurassic-2 (Text) & ISR & Closed & [\href{https://docs.ai21.com/docs/responsible-use}{4}] \\
AI2 & AI2 ImpACT License for Low-Risk Artifacts & Y (Tulu v2) & License & OLMo (Text) & USA & Open & [\href{https://allenai.org/licenses/impact-lr}{5}] \\
Aleph Alpha & Terms and Conditions, §4.8 Customer’s Rights and Use Restrictions & N & TOS & Luminous (Text) & DEU & Closed*** & [\href{https://aleph-alpha.com/terms-conditions/}{6}] \\
Amazon & AWS Responsible AI Policy \& \newline AWS Acceptable Use Policy* & N & Standalone & Titan Text (Text) & USA & Closed & [\href{https://aws.amazon.com/machine-learning/responsible-ai/policy/}{7}] \\
Anthropic & Acceptable Use Policy & N & Standalone & Claude 3 (Text) & USA & Closed & [\href{https://www.anthropic.com/legal/archive/4903a61b-037c-4293-9996-88eb1908f0b2}{8}] \\
Baidu & ERNIE Bot User Agreement, §4 Service Usage Specifications & Y (ERNIE) & TOS & ERNIE 4.0 (Text) & PRC & Closed & [\href{https://yiyan.baidu.com/infoUser}{9}] \\
BigCode & BigCode Open RAIL-M v1 License, \newline §A Use Restrictions & Y (StarCoder 2) & License & StarCoder 2 (Text) & N/A** & Open & [\href{https://huggingface.co/spaces/bigcode/bigcode-model-license-agreement}{10}] \\
BigScience & BigScience RAIL License v1.0, \newline §A Use Restrictions & Y (BLOOM) & License & BLOOM (Text) & N/A** & Open & [\href{https://huggingface.co/spaces/bigscience/license}{11}] \\
Character.AI & Terms of Service, Conditions of Use & N & TOS & Not Public (Text) & USA & Closed & [\href{https://beta.character.ai/tos}{12}] \\
Cohere & Usage Guidelines & N & Standalone & Command (Text) & CAN & Closed & [\href{https://docs.cohere.com/docs/usage-guidelines}{13}] \\
Databricks & Databricks Open Model Acceptable Use Policy & Y (DBRX) & Standalone & DBRX (Text) & USA & Open & [\href{https://www.databricks.com/legal/acceptable-use-policy-open-model}{14}] \\
DeepSeek & Terms of Use, §3 Service Management & N & TOS & DeepSeek (Text) & PRC & Open & [\href{https://chat.deepseek.com/downloads/DeepSeek\%20Terms\%20of\%20Use.html}{15}] \\
Eleven Labs & Terms of Service, Prohibited Activities & N & TOS & Not Public**** (Audio) & USA & Closed & [\href{https://elevenlabs.io/terms}{16}] \\
Google & Generative AI Prohibited Use Policy & N & Standalone & Gemini (Multimodal) & USA & Closed & [\href{https://policies.google.com/terms/generative-ai/use-policy}{17}] \\
Inflection & Terms of Service, Acceptable Use & N & TOS & Inflection-2.5 (Text) & USA & Closed & [\href{https://pi.ai/policy\#terms}{18}] \\
Meta & Acceptable Use Policy & Y (Llama 2) & License & Llama 2 (Text) & USA & Open & [\href{https://llama.meta.com/llama3/use-policy/}{19}] \\
Midjourney & Terms of Service, §9 Community Guidelines & N & TOS & Midjourney v6 (Image) & USA & Closed & [\href{https://docs.midjourney.com/docs/terms-of-service}{20}] \\
Mistral & Terms of Use, §8 Your obligations/§9 Our Obligations \& Le Chat Terms of Service, §4.3 Chat Moderation Policy* & Y (Mistral API) & TOS & Mixtral (Text) & FRA & Open & [\href{https://mistral.ai/terms/\#terms-of-use}{21}] \\
OpenAI & Usage Policy & N & Standalone & GPT-4 (Multimodal) & USA & Closed & [\href{https://openai.com/policies/usage-policies/}{22}] \\
Perplexity & Terms of Service, Acceptable Use & N & TOS & Not Public**** (Text) & USA & Closed & [\href{www.perplexity.ai/hub/legal/terms-of-service}{23}] \\
Reka & Terms of Service, §3.2 Responsible Use & N & TOS & Yasa-1 (Multimodal) & USA & Closed & [\href{www.reka.ai/terms-of-use/}{24}] \\
Runway & Terms of Service, §5 User Conduct & N & TOS & Not Public**** (Video) & USA & Closed & [\href{runwayml.com/terms-of-use/}{25}] \\
Stability AI & Acceptable Use Policy & N & Standalone & Stable Diffusion 3 (Image) & GBR & Open & [\href{stability.ai/use-policy}{26}] \\
TII & Acceptable Use Policy & Y (Falcon 180B) & Standalone & Falcon 180B (Text) & UAE & Open & [\href{falconllm.tii.ae/acceptable-use-policy.html}{27}] \\
Together & Terms of Service, §2.4 Your Responsibilities & N & TOS & StripedHyena Nous (Text) & USA & Open & [\href{www.together.ai/terms-of-service}{28}] \\
Twelve Labs & Terms of Service, §14 No Unlawful or Prohibited Use & N & TOS & Pegasus-1 (Video) & USA & Closed & [\href{docs.google.com/document/d/18-n27z3V98Kz-57EIozg-gYmtIfdlrykvP-ciyTm17M/edit}{29}] \\
Writer & Terms and Conditions, §4.3 Acceptable Use & N & TOS & Palmyra-1 (Text) & USA & Closed & [\href{writer.com/legal/terms-of-use/}{30}] \\
\bottomrule
\end{tabular}}
\caption{\textbf{Foundation Model Developers' Acceptable Use Policies.} This table includes information on the 30 acceptable use policies under consideration in this paper, including: the developer's name; the title of the developer's AUP and (if applicable) the section of that policy that includes use restrictions; whether the policy as applied by the developer is specific to a certain foundation model (and if so which foundation model); the type of policy document that contains the AUP (a model license, a terms of service agreement, or a standalone policy); the developer's flagship foundation model series (and the output modality of those models); the country in which the developer is headquartered; whether the weights of the flagship model series are open; and a reference to the AUP. *Amazon and Mistral's TOS explicitly refer to two relevant documents, so both are considered. **International research coalitions. ***Aleph Alpha provides model weights to customers on premises. ****These developers have not publicly disclosed the name of or details about their flagship foundation models. (Last updated April 18, 2024)}
\label{tab:aups}
\end{table*}
\needspace{1\baselineskip}

Neither the EU AI Act nor the US Voluntary AI Commitments require that firms enforce their AUPs or restrict any particular uses.
By contrast, China's February 2024 regulatory guidance on Basic Safety Requirements for Generative Artificial Intelligence Services specifies 31 safety risks that developers must prohibit, such as ``subvert[ing] state power,'' ``endanger[ing] national security,'' and ``dissemination of false and harmful information'' \citep{China_Safety_AI_2024, zeng2024airiskcategorizationdecoded}.

\section{3. Methodology}
\textbf{3.1 Search protocol for acceptable use policies} \newline 
\noindent Table \ref{tab:aups} details 30 foundation model developers' acceptable use policies. Developers use different policy documents to limit model use, including: a standalone acceptable use policy for all their foundation models (e.g., Google, Stability AI), use restrictions included in a general model license (e.g., AI2), use restrictions included in a custom model license (e.g., BigScience, Meta), or provisions in terms of service agreements that apply to all services including foundation models (e.g., Midjourney, Perplexity, Eleven Labs). 

The following protocol was used to identify acceptable use policies across these different types of documents:
\begin{enumerate}[itemsep=0pt,parsep=0pt,topsep=0pt]
    \item Compile a list of foundation model developers using the data provided by \cite{bommasani2023ecosystem}.
    \item For each developer, check the terms of service (TOS) on its website. If the TOS include an AUP with content restrictions that plausibly cover the developer's foundation models, take that portion of the TOS as the AUP.
    \item For each remaining developer, check the license for its “flagship foundation model'';\footnote{A flagship foundation model is a developer's most salient and/or capable model, informed by its public documentation \citep{bommasani2023foundation}.} if it includes behavioral use restrictions, take that portion of the license as the AUP.
    \item For each remaining developer, if the TOS or license reference a separate document with behavioral use restrictions (e.g., usage guidelines) such that the restrictions are binding, take the relevant portion of that document as the AUP. \newline
\end{enumerate}

\noindent \textbf{3.2 Coding of prohibited use categories in AUPs} \newline
%shifting blame to consumers
\noindent Qualitative content analysis was used for this paper's coding of prohibited use categories in developers' acceptable use policies \citep{Mayring2015}. 
This was done inductively \citep{https://doi.org/10.1111/j.1365-2648.2007.04569.x}, with categories drawn directly from acceptable use policies, and was inspired by prior work related to AI ethics guidelines \citep{Jobin2019, Fjeld2020}, privacy policies \citep{Alfawzan2022}, content moderation guidelines \citep{election2021longfuse}, benchmarks \citep{wang2024decodingtrust}, and Responsible AI Licenses \citep{mcduff2024standardization}.
%add another RAIL cite

The following process was used to code the prohibited use categories included in developers' acceptable use policies: 
\begin{itemize}[itemsep=0pt,parsep=0pt,topsep=0pt]
    \item For each acceptable use policy, each line of the policy was analyzed. For each line, the distinct prohibited use categories included were added to a list of prohibited uses across every developers' acceptable use policy. Distinct prohibited use categories do not include different types of actions related to the same prohibited use category (e.g., ``generating, promoting, or further distributing spam'' was coded as ``spam'') or categories with substantial overlap that do not use distinct phrasing. 
    \item Using the list of prohibited use categories across all AUPs, each line of each acceptable use policy was considered again to ensure the prohibited use categories therein are coded correctly. A prohibited use category should receive a specific coding only if it uses near-identical language to that coding, and each prohibited use category in each policy receives only one coding.
\end{itemize}

\noindent This produced a list of 127 categories and a 30x127 matrix (visible on \href{https://github.com/kklyman/aupsforfms}{GitHub}), where columns show foundation model developers, rows show prohibited use categories, and cells are marked ``1'' if a developer's acceptable use policy explicitly references that prohibited use category and ``0'' otherwise. 
Section 4 analyzes the results of this coding.

This methodology satisfies three aims. 
First, it provides a systematic and comprehensive approach for capturing the prohibited use categories included in acceptable use policies.
Second, it enables a granular analysis of acceptable use policies.
Classifying prohibited use categories into higher-level groups is an illustrative exercise (see Figure \ref{fig:counts}), but acceptable use policies are legal documents with unique provisions that require close study \citep{doi:https://doi.org/10.1002/9781118978238.ieml0001}.
Third, it clarifies the risks from foundation models that developers themselves seek to mitigate.
While many previous works have taxonomized the risks and harms stemming from foundation models \citep{weidinger2023sociotechnical, shelby2023sociotechnical, Hoffmann2023, Fergusson2023, atkinson2024legalrisktaxonomygenerative, marchal2024generativeaimisusetaxonomy}, this paper assesses how companies taxonomize risk on the basis of their own policies. 

\section{4. Analysis of acceptable use policies}

\textbf{4.1 Developers with acceptable use policies}

\noindent Foundation model developers that have AUPs are heterogeneous along multiple axes, demonstrating broad adoption (see Table \ref{tab:aups}). 
In terms of model release, 12 of the developers openly release the model weights for their flagship model series, while 18 do not. 
These models have a variety of different output modalities, with 20 language models, 4 multimodal models, 3 image models, 2 video models and 1 audio model. 
The developers are headquartered around the world, with 19 based in the US and the others based in Canada, China, France, Germany, Israel, and the UAE. \newline

\begin{figure}[h!]
    \centering
    \includegraphics[width=\linewidth]{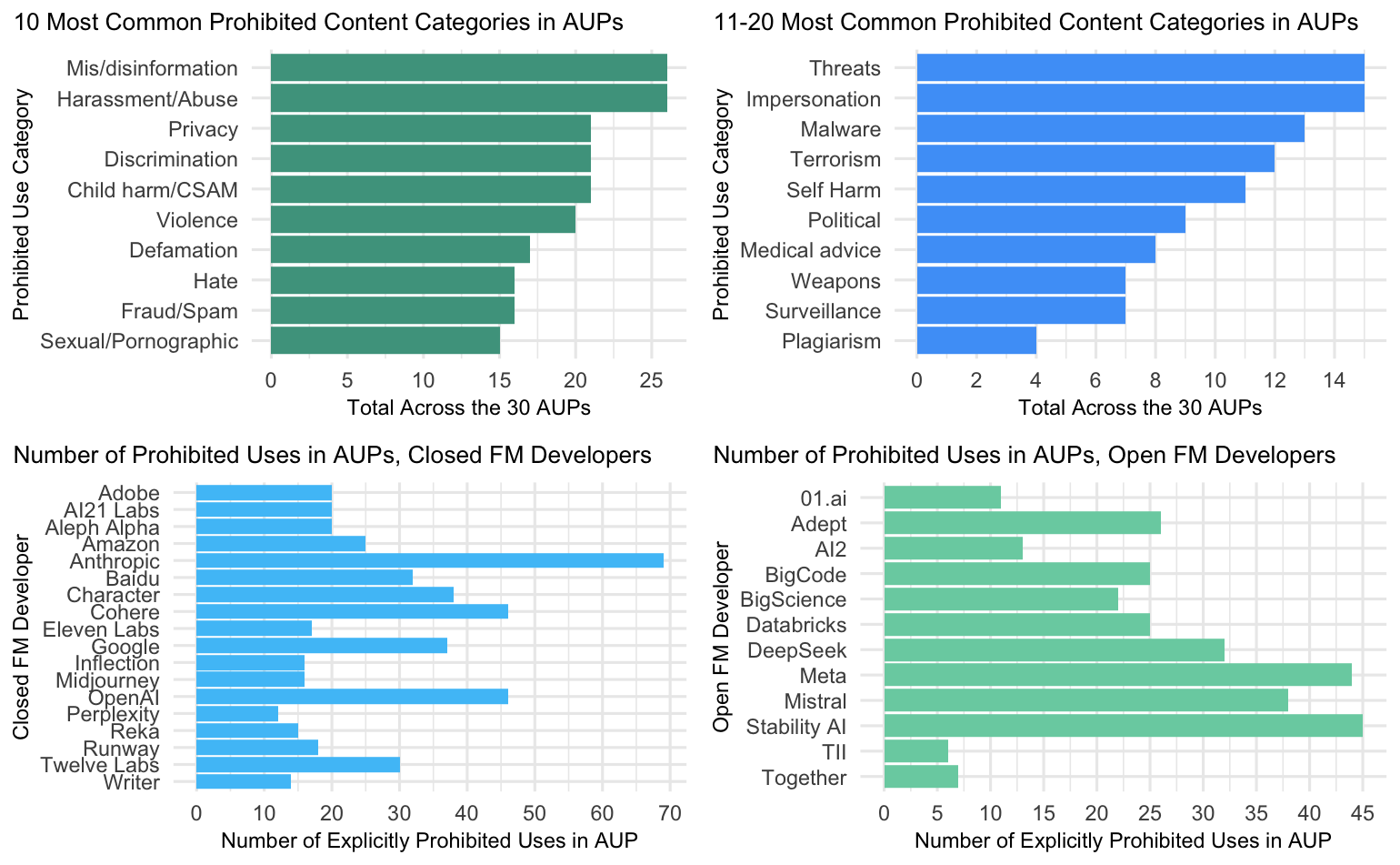}
    \captionsetup{skip=2pt}
    \caption{\textbf{Common prohibited content categories and number of prohibited uses per developer.} Top left: the 10 most common categories of content-related prohibited uses in developers' AUPs. Top right: the next 10 most common categories of content-related prohibited uses in developers' AUPs. (See the \href{https://github.com/kklyman/aupsforfms/tree/main}{GitHub} for details on grouping.) Bottom left: the number of explicitly prohibited uses in closed developers' AUPs (out of 127 categories). Bottom right: the number of explicitly prohibited uses in open developers' AUPs.}
    \label{fig:counts}
\end{figure}

\noindent \textbf{4.2 Prohibited content in acceptable use policies} \newline
Acceptable use policies commonly prohibit users from employing foundation models to generate content that is explicit (e.g., violence, pornography), fraudulent (scams, spam), abusive (harassment, hate speech), deceptive (disinformation, impersonation), or otherwise harmful (malware, privacy infringements).\footnote{Content-based restrictions generally apply only to user prompts that request that a model generate this type of content---models will classify the toxicity of this type of content if asked to do so, but it is against developers' policies to generate such content.}
Figure \ref{fig:counts} shows the most common categories of content that are explicitly prohibited by developers' acceptable use policies: mis/disinformation (26 policies include explicit prohibitions), harassment/abuse (26), privacy (21), discrimination (21), and child harm/child sexual abuse material (21) were the most frequent, while categories like political content (9), medical advice (8), weapons (7), surveillance (7), and plagiarism (4) were less common. \newline
\indent Many developers' acceptable use policies have granular use restrictions, whereas others have broad restrictions without much elaboration. 
Figure \ref{fig:counts} shows the number of prohibited use categories contained in each developers' acceptable use policy and distinguishes between open- and closed-weight developers \citep{kapoor2024societal}.
Among closed developers, the acceptable use policies of Anthropic (69 prohibited uses), Cohere (46), and OpenAI (46) explicitly reference the largest number of prohibited use categories, while the policies of smaller startups such as Reka (15), Writer (14), and  Perplexity (12) have the fewest. 
Among open developers, the acceptable use policies of Stability AI (44), Meta (44), and Mistral (38) explicitly reference the largest number of prohibited use categories, while the AUPs of 01.ai (11), Together (7), and the Technology Innovation Institute (6) reference the fewest. 
The average number of prohibited uses for closed developers is 20 (standard deviation of 15.1), while the average for open developers is 24.5 (standard deviation is 13.5). \newline 
\indent There are several potential explanations for open developers having a larger number of prohibited use categories in their AUPs. 
Open foundation model developers often use Responsible AI Licenses that feature a sizable, standardized set of use restrictions \citep{Contractor_2022, Keller2023Growth}. 
Second, a greater number of closed foundation model developers have acceptable use policies (including smaller companies without large legal teams), whereas many other open developers have no acceptable use policy (see Table \ref{tab:noaups}), introducing potential selection bias in computing the average. 
Third, unlike closed developers, open developers often cannot enforce their acceptable use policies against individual users, so prohibiting a larger number of uses may come at less cost. \newline  %ADD REFAPP
\indent The strength of an acceptable use policy is not determined solely by the number of prohibited uses it lists.
All 30 acceptable use policies prohibit users from generating content that violates the law, and the majority prohibit users from generating content that impedes the model developer’s operations or is not accompanied by adequate disclosure that it is machine-generated. 
These catch-all prohibitions cover unenumerated risk categories, making acceptable use policies more malleable and comprehensive by linking them to laws and organizational procedures that may change. 
Over 40 of the 127 prohibited use categories relate to potentially illegal content (e.g., child sexual abuse material, defamation, discrimination against a protected class, drugs, fraud, hate speech, malware, prostitution, scams), reflecting the fact that developers consider these to be risks associated with their models and wish to reduce their liability for such risks \citep{Lemley2023}.

\vspace{-3pt}
\vspace{-4pt}

\indent Prohibitions on content that is not generally illegal show developers' priorities and highlight different approaches taken in their acceptable use policies. 
Political content, such as using foundation models for campaigning, lobbying, or otherwise influencing political processes, is explicitly prohibited by 9 startups---Anthropic, Character.AI, Cohere, Databricks, Midjourney, OpenAI, Perplexity, Stability AI, and Twelve Labs---whereas Big Tech companies like Amazon, Google, and Meta have no such prohibitions.  
Weapons-related content is explicitly prohibited by 7 developers: AI2, Anthropic, Amazon, Meta, Mistral, OpenAI, and Stability AI. 
Generating eating disorder-related content, such as pro-anorexia content, is explicitly prohibited by just 4 developers: Character.AI, Cohere, Meta, and Mistral.
And while some open developers such as Adept, DeepSeek, and Together broadly prohibit some types of sexual content, others like Meta and Mistral prohibit only content related to prostitution or sexual violence.
Foundation models have the potential to cause harm in each of these areas, yet major developers choose not to adopt legally binding restrictions on using their models in these ways \citep{10.1093/pnasnexus/pgae034, doi:10.1080/21624887.2020.1760587, Sharp2023}.

Other notable prohibited uses include:
\begin{itemize}[itemsep=0pt,parsep=0pt,topsep=0pt]
    \item \textit{Undermining the interests of the state}: Baidu and DeepSeek, two of three model developers in Table \ref{tab:aups} headquartered in China, state in their acceptable use policies that users must not generate content “endangering national security, leaking state secrets, subverting state power, overthrowing the socialist system, and undermining national unity…damaging the honor and interests of the state…undermining the state's religious policy”. 01.ai, the other Chinese developer, also includes a prohibition against “harming national security.” These restrictions draw directly on China’s Basic Safety Requirements for Generative AI Services \citep{zeng2024airiskcategorizationdecoded}.
    \item \textit{Password trafficking}: Eleven Labs, the only developer in Table \ref{tab:aups} whose flagship model outputs audio, prohibits users from using its models to “trick or mislead us or other users, especially in an attempt to learn sensitive account information, for example user passwords.” This may be intended to address concerns regarding the use of voice cloning for scams \citep{marchal2024generativeaimisusetaxonomy, Barnett_2023}. 
    \item \textit{Misinformation}: The extent to which developers restrict users’ ability to generate and/or distribute inaccurate content varies widely. While some AUPs include wholesale bans on misinformation (e.g., AI21 Labs, Inflection), others have looser restrictions that apply only to verifiable disinformation with the intent to cause harm (e.g., TII). Mis/disinformation is the most frequently prohibited category of use---even more so than child sexual abuse material---indicating that some developers may be more responsive to political and reputational risk than assessments of harm or legal liability \citep{pfefferkorn_2024, thiel_stroebel_portnoff_2023, geng2023comparing}.  \\
\end{itemize}

\noindent \textbf{4.3 Restrictions on types of end use}

\noindent In addition to content-based restrictions, acceptable use policies for foundation models often restrict the types of activity that users can carry out. %engage in when using their models
Acceptable use policies from 6 developers prohibit ``model scraping'' or training a model on their own model's outputs. 
Anthropic’s Acceptable Use Policy bans use of “prompts and results to train an AI model (e.g., ‘model scraping’)”; Adept, Adobe, Meta, Perplexity, and Runway similarly prohibit the use of model outputs for training other foundation models.
While 8 developers have no such explicit ban (BigCode, BigScience, Character.AI, Eleven Labs, Mistral, Stability AI, TII, Reka), the remaining 16 prohibit the use of their models to build a competing service, which encompasses model scraping \citep{metz2024adobe}.

Some developers prohibit using their models to distribute AI-generated content at scale. 
AI21 Labs’ Usage Guidelines state that “No content generated by AI21 Studio will be posted automatically (without human intervention) to any public website or platform where it may be viewed by an audience greater than 100 people.” 
Four other developers (BigCode, BigScience, Cohere, and Databricks) prohibit using their models for automated posting online \citep{goldstein2023generativelanguagemodelsautomated}.  

Many acceptable use policies prevent firms in certain industries from making use of foundation models. 
For example, weapons manufacturers would be in violation of a policy with weapons-related restrictions if they made use of the foundation model to produce weapons, though it is possible that the developer negotiates custom contracts with weapons manufacturers \citep{brenes2024private, simmonsedler2024aipoweredautonomousweaponsrisk}. 
In January 2024, OpenAI reportedly changed its Usage Policies to facilitate partnerships with militaries, deleting a line that prohibited use related to ``military and warfare" \citep{biddle2024openai, schatz2024letter}.

\begin{table*}[ht!]
\resizebox{\textwidth}{!}{
\begin{tabular}{p{2cm}lp{16cm}p{2cm}p{3cm}}
\toprule
\textbf{Developer} & \textbf{Model} & \textbf{Intended Use (Source)} & \textbf{Model Assets Released} & \textbf{License} \\
\midrule
Alibaba Cloud & Qwen-VL & “Researchers and developers are free to use the codes and model weights of both Qwen-VL and Qwen-VL-Chat. We also allow their commercial use.” (\href{https://github.com/QwenLM/Qwen-VL/tree/master}{License Blurb}) & Code, weights & Tongyi Qianwen \newline License Agreement \\
EleutherAI & GPT-NeoX 20B & “Developed primarily for research purposes…. not intended for deployment as-is. It is not a product and cannot be used for human-facing interactions without supervision.” (\href{https://huggingface.co/EleutherAI/gpt-neox-20b}{Model Card}) & Data, \newline code, weights & Apache 2.0 \\
Meta & MusicGen-Large & “The model should not be used on downstream applications without further risk evaluation and mitigation. The model should not be used to intentionally create or disseminate music pieces that create hostile or alienating environments for people. This includes generating music that people would foreseeably find disturbing, distressing, or offensive; or content that propagates historical or current stereotypes.” (\href{https://huggingface.co/facebook/musicgen-large}{Model Card}) & Data, \newline code, weights & CC-BY-NC 4.0 \\
Microsoft & Phi-2 & “Given the nature of the training data, the Phi-2 model is best suited for prompts using the QA format, the chat format, and the code format. … Direct adoption for production tasks without evaluation is out of scope of this project.” (\href{https://huggingface.co/microsoft/phi-2}{Model Card}) & Weights & MIT \\
Mistral & Mixtral-8x7B & N/A (\href{https://huggingface.co/mistralai/Mixtral-8x7B-v0.1}{Model Card}) & Code, weights & Apache 2.0 \\
Databricks & MPT-30B & “Not intended for deployment without finetuning. It should not be used for human-facing interactions without further guardrails and user consent.” (\href{https://huggingface.co/mosaicml/mpt-30b}{Model Card}) & Code, weights & Apache 2.0 \\
xAI & Grok-1 & “Grok-1 is intended to be used as the engine behind Grok for natural language processing tasks including question answering, information retrieval, creative writing and coding assistance.” (\href{https://x.ai/blog/grok/model-card}{Model Card}) & Weights & Apache 2.0 \\
\bottomrule
\end{tabular}}
\caption{\textbf{Foundation Model Developers Without Acceptable Use Policies.} Information on developers without acceptable use policies, including the name of the developer, the name of the model where no acceptable use policy has been applied, the intended use of that model (and the source), the model assets released, and the license under which the model is distributed.}
\label{tab:noaups}
\end{table*}
%\textbf{ADD REF}

Acceptable use policies may also restrict the use of models in highly-regulated industries such as law, finance, and medicine. 
8 of the 30 acceptable use policies include restrictions on medical advice, and Anthropic, Character.AI, Google, Meta, and OpenAI also have restrictions on legal and financial advice, which apply not only to lawyers, doctors, and financial advisers, but also to organizations that provide services in these fields \citep{doi:10.1056/NEJMhle2308901, Dahl_2024, tamkin2023evaluatingmitigatingdiscriminationlanguage}. \newline 
\indent AI2, Amazon, Anthropic, Google, and OpenAI also prohibit use of their models for certain types of surveillance. 
Google prohibits use of its models for “tracking or monitoring people without their consent” while AI2 singles out “military surveillance.”
This could prevent spyware companies and defense and intelligence contractors respectively from making use of their foundation models \citep{feldstein2019global, 10.1145/3372823}.

\vspace{-2pt}
\begin{figure}[h!]
    \centering
    \includegraphics[width=0.8\linewidth]{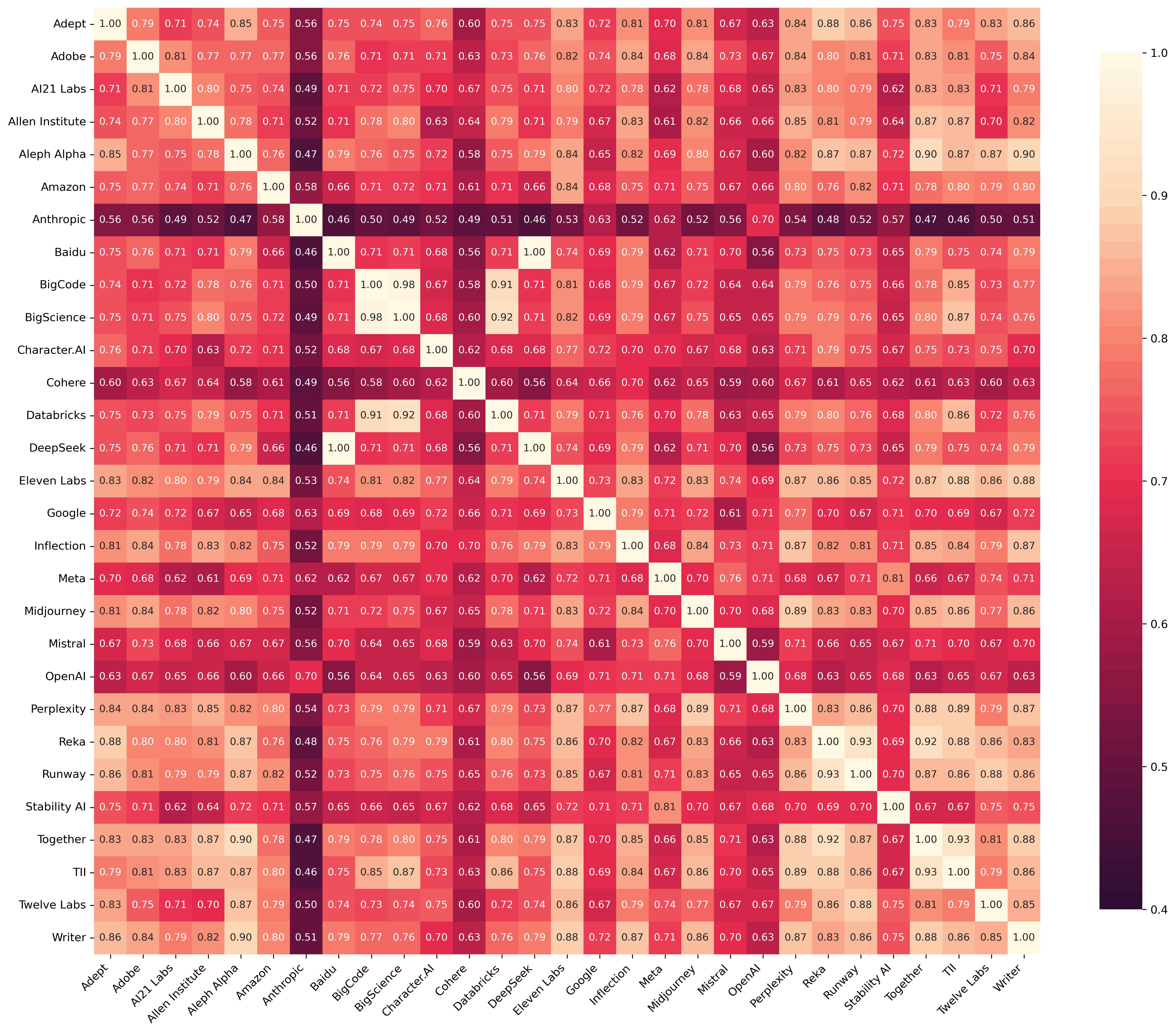}
    \captionsetup{skip=1pt}
    \caption{\textbf{Developer correlations.} The correlation between prohibited use categories for pairs of developers across all 127 categories. Correlation is measured using the simple matching coefficient (i.e. agreement rate), which is the fraction of all indicators for which both developers are assigned the same value (i.e. where both are assigned 1 as both of their AUPs prohibit the category, or both are assigned 0). 
    \label{fig:corrs}}
\end{figure}

\noindent \textbf{4.4 Correlations between developers' AUPs} \newline
\noindent Despite increased standardization across open developers \citep{mcduff2024standardization}, acceptable use policies remain inconsistent across foundation model developers. 
Figure \ref{fig:corrs} shows the correlation between developers’ acceptable use policies based on which of the 127 prohibited use categories they include. 
BigCode, BigScience, and Databricks have highly similar policies (with a correlation of more than 0.9), as do Baidu and DeepSeek (the two Chinese developers) and Reka, TII, and Together (developers with relatively few prohibited use categories).
Anthropic, Cohere, Google, Meta, Mistral, OpenAI, and Stability AI are among the developers with the policies that are least similar to others, in part because they have the largest number of prohibited uses; each have a correlation of 0.7 or less with 15 or more other developers. 
This may pose an issue for cloud providers that distribute models from many developers; Amazon Web Services, for example, distributes models from AI21 Labs, Amazon, Anthropic, Cohere, Meta, Mistral, and Stability AI, but Anthropic's acceptable use policy has a correlation of less than 0.6 with those each of these other developers, indicating AWS would need to enforce several substantially different policies.\\

\noindent \textbf{4.5 Developers without acceptable use policies} \newline
\noindent There are tens of developers 
that do not have acceptable use policies for their foundation models---table \ref{tab:noaups} provides seven examples.
There are myriad reasons why a developer may choose to release a model without acceptable use policy. 
Some open foundation model developers do not use acceptable use policies because their models are intended for research purposes only---if they were to adopt use restrictions, it could deter researchers from conducting safety research through red teaming or adversarial attacks (in the absence of a safe harbor for good faith researchers) \citep{basdevant2024frameworkopennessfoundationmodels, longpre2024safe}. 
Other models intended for research may lack acceptable use policies on the basis that they present less severe risks of misuse, whether because they have less significant capabilities or fewer users \citep{eiras2024risksopportunitiesopensourcegenerative}. 
Non-commercial models such as these are frequently distributed using licenses without use restrictions such as Apache 2.0 or Creative Commons Attribution-NonCommercial licenses \citep{white2024modelopennessframeworkpromoting, longpre2024responsiblefoundationmodeldevelopment}. 
While a license may not include any use restrictions for noncommercial users, commercial users may have to agree to custom use restrictions in their contracts with the model developer, which are not public.
This creates a potential information asymmetry where a developer and its clients are aware of the domains in which use is permitted, while regulators and the public may be led to believe that model use is unrestricted \citep{hacker2024regulating, truby2022sandbox}.

Foundation models available for commercial use may not include acceptable use policies for several reasons. 
In some cases, developers offer a model ``as is,'' stating that it is not intended for commercial use without further fine-tuning, mitigations, or use restrictions by downstream developers (e.g., Databricks’ MPT-30B). 
Developers hoping to maximize uptake among commercial users may be less likely to adopt acceptable use policies because clients' risk-averse legal teams could recommend using different models without such restrictions.  
Other developers release their models without complete documentation, whether because they intend to release an acceptable use policy at a later point, which could be part of staged release, or due to under-documentation in the rush to release a model \citep{10.1145/3593013.3593981, mitchell2023pillars}.  \\
\indent In any case, other restrictions may apply to foundation models without acceptable use policies. 
Alibaba Cloud restricts firms with over 100 million users from making use of Qwen-VL through its license, which also bans model scraping. 
Restrictions on who can use a foundation model may have a significant effect on how it is used even in the absence of legally binding behavioral use restrictions \citep{bommasani2024foundationmodeltransparencyindex}.

\section{5. Enforcement of acceptable use policies}

\noindent \textbf{5.1 Barriers to enforcement}

\textit{5.1.1 Practical and legal barriers for open developers}

\noindent The enforceability of open foundation model developers' acceptable use policies is a major limitation on how effective they are at restricting risky uses. 
Unlike closed foundation model developers, whose models are distributed via their own products, services, or APIs (or those of another firm), developers of open foundation models distribute their models by distributing the weights online such that they can be downloaded, and models are often run locally \citep{10.1145/3593013.3593981}. 
As a result, open developers have few ways of monitoring downstream use of their models, making it difficult for them to enforce their policies where models are run locally or where hosted inference is provided by another organization \citep{srikumar2024risk}. \newline
\indent If open foundation model developers were to attempt to enforce their acceptable use policies, many would face substantial legal barriers. 
Licenses for open-weight foundation models that include behavioral use restrictions are a type of copyright license, but it is unclear if machine learning models are copyrightable artifacts, calling into question the enforceability of such licenses \citep{henderson2023foundation, lee2024talkin}. 
\cite{downing2023licensing} argues that even if Responsible AI Licenses for models do not trigger copyright issues, the use restrictions in these licenses are ineffective as licensees are not required to enforce them against downstream licensees and developers themselves cannot sue downstream licensees for violations. Licenses for open-weight models also face issues related to interoperability, as use restrictions may not propagate to software that receives inputs from the model \citep{downing2024choose}. \newline
\indent On the other hand, private sector licensees will likely comply with acceptable use policies of open-weight foundation models due to the legal risk associated with noncompliance. 
In cases where an open developer does not seek to enforce its acceptable use policy, the policy can still encourage responsible use \citep{10.1145/3593013.3594002}. 
Most users are not bad actors and may adhere to a policy despite gaps in enforcement, as they have no interest in generating prohibited content. \newline %reCheck Dave's edit 
\indent Despite these challenges, many open foundation model developers attempt to restrict the use of their models to some degree. 
12 of the developers that have acceptable use policies openly release their flagship model's weights, but do so using licenses or terms of service that prohibit certain unacceptable uses. 
Although open foundation models are frequently referred to as “open-source” in popular media, truly open-source software or machine learning models cannot have use restrictions by definition \citep{OSI2024OpenSourceAI, downing2024choose}. \newline
\indent \textit{5.1.2 Ecosystem barriers} \newline
\noindent Another issue in gauging the enforcement of acceptable use policies is the way in which they propagate across the foundation model ecosystems. 
In addition to developers, cloud service providers (e.g., AWS, Azure, GCP) and other digital platforms (e.g., Salesforce, Scale AI) act as deployers of foundation models that were not developed in-house. 
Deployers have their own acceptable use policies for their platforms that do not align perfectly with external developers’ acceptable use policies, and it is not clear that a deployer would have adequate expertise to restrict the uses of a foundation model in accordance with an acceptable use policy that is more stringent than that of the deployer \citep{https://doi.org/10.1002/itl2.84}.
In particular, deployers would need to build infrastructure to support enforcement of the distinct acceptable use policies for each of the foundation models they distribute.
While there are a variety of publicly available models and tools that deployers might leverage to enforce developers' acceptable use policies (e.g., by filtering specific categories of prompts and responses), there is little evidence deployers have done so. \newline
\indent As an alternative, a deployer may attempt to devise (and enforce) its own acceptable use policy that encompasses those of each of its developer partners. 
However, the large variation in prohibited use categories among different developers' acceptable use policies makes such an exercise difficult, and would require that for each category the deployer apply the most restrictive of its partners' acceptable use policies to every model.
\cite{gorwa2024moderating} find that model marketplaces such as Hugging Face and GitHub have struggled to enforce their own acceptable use policies in light of the challenge of moderating the distribution of thousands of machine learning models, each of which may come with its own use restrictions \citep{huggingface2023content, github2024acceptable}. \\
\indent These challenges are made more stark by the ease with which users can circumvent technical measures used to enforce acceptable use policies.
\cite{zeng2024airbench2024safetybenchmark} show that including uncommon dialects and appeals to authority in prompts can cause a foundation model to violate its developer's acceptable use policy despite safety filters in APIs.
In addition, \cite{qi2023finetuning} find that fine-tuning foundation models via an API can remove safety measures like instruction tuning and RLHF such that models will more readily violate their developer's acceptable use policy.
%give more quant findings
Other researchers have found many vulnerabilities that allow users to nullify measures intended to promote adherence to acceptable use policies, such as adversarial prompts \citep{maus2023black,robey2023smoothllm}, jailbreaks \citep{wei2023jailbroken, shen2023anything, zou2023universal, shah2023scalable}, and other methods for fine-tuning away safety measures via APIs \citep{yang2023shadow, zhan2023removing}. 
These vulnerabilities show that closed developers are likely unable to enforce their acceptable use policies in many cases \citep{henderson2024safety}.

\textit{5.1.3 Misallocating responsibility to users}

\noindent Acceptable use policies are a means of shifting responsibility (and liability) for risky uses of a technology from the developer, deployer, or distributor of that technology to the user \citep{doherty2011reinforcing, WeidmanGrossklags2019}. 
Acceptable use policies may be effective in limiting the behavior of corporate users, which are legally risk-averse, but are unlikely to fundamentally change the behavior of the average individual user \citep{villa2022evaluating}.

Developers' approach to indemnification crystallizes the issue. Meta's Llama licenses, for example, hold users responsible for any direct or downstream use of the model, stating ``[y]ou will indemnify and hold harmless Meta from and against any claim by any third party arising out of or related to your use or distribution of the Llama Materials.''

Social media companies' content policies also shift responsibility for toxic content from the platform that algorithmically amplifies such content to individual users that post it \citep{keller2021amplification}.
The same can be said of AI ethics guidelines, which often provide guidance to users regarding how to ethically use a company's AI systems rather than describing the tangible steps a company will take to prioritize ethics above other aims \citep{10.1145/3461702.3462622, Fjeld2020}.
Similarly, developers employ acceptable use policies to eschew responsibility for downstream impacts of the foundation models they choose to build and deploy.

Acceptable use policies often impose obligations on users that they are ill-equipped to uphold. 
Setting aside issues of digital literacy \citep{NG2021100041}, the user is often not the right party to be responsible for ensuring that a foundation model is not generating violative outputs \citep{10.1145/3531146.3533150}. 
For instance, holding users responsible for generating self-harm related content may be viable for users that maliciously seek to spread such content online, but not for vulnerable users seeking to harm themselves and who turn to a foundation model for aid \citep{Grabb2024.04.07.24305462}. 

One solution that developers implement is increasing surveillance of their users to monitor dangerous prompts and responses. 
\cite{Robinson2019} argues surveillance is a fundamental feature of acceptable use policies, as they are leveraged by powerful institutions as a mode of control over their subjects.
Enforcing acceptable use policies often requires developers to monitor users' interactions with foundation models closely, which could facilitate privacy breaches if data protection is inadequate \citep{wachter2019right, kayes2017privacy}.   \\
 
\noindent \textbf{5.2 Potential negative externalities of enforcement}

\textit{5.2.1 Restricting researcher access}

\noindent \cite{longpre2024safe} find that of seven major foundation model developers with acceptable use policies, none provide comprehensive exemptions for researchers. 
Platforms that distribute foundation models may rate limit or ban accounts that violate acceptable use policies, even if those accounts belong to researchers, meaning that acceptable use policies can act as a disincentive against carrying out adversarial red teaming.
Concerns regarding restrictions on researcher access led over 350 researchers and advocates to sign an open \href{https://sites.mit.edu/ai-safe-harbor/}{letter} calling for companies to refrain from disproportionate enforcement of their acceptable use policies in such cases.

\textit{5.2.2 Case studies of AUPs preventing beneficial uses}

\noindent Strict acceptable use policies can inadvertently prevent a wide variety of beneficial uses of foundation models. Acceptable use policies do not permit users to generate prohibited content when doing so would likely be net beneficial in a specific context or circumstance, meaning that they function as a blanket ban on certain types of content \citep{small2023black}.

Acceptable use policies can ban entire domains of use, but this might be overly cautious in scoping out applications.  
For example, Meta's acceptable use policy for Llama 2 states ``You agree you will not use, or allow others to use, Llama 2 to...Engage in, promote, incite, facilitate, or assist in the planning or development of activities that present a risk of death or bodily harm to individuals, including use of Llama 2 related to...Operation of critical infrastructure, transportation technologies, or heavy machinery''.
Critical infrastructure and heavy machinery are not defined in the policy, making this restriction expansive. 
If a robotics company were to use Llama 2 to assist in turning transcribed audio instructions into commands for a robot, 
Meta would plausibly have a claim that the company had violated its prohibition against using Llama 2 to assist with heavy machinery.
Language models are used in numerous ways in robotics research, and acceptable use policies could limit such research \citep{brohan2023rt2visionlanguageactionmodelstransfer, kim2024openvlaopensourcevisionlanguageactionmodel}. 

Acceptable use policies can also prevent the use of models to generate content that developers consider obscene, even when it could be beneficial \citep{stardust2018safe, jones2020camming, Bronstein2021deplatforming}. 
In preventing generation of sexual content, an acceptable use policy would prohibit the use of a foundation model to assist in reducing harm associated with sex work \citep{bernier2021use, sanders2018internet, rekart2005sex}. 
Sex workers would be prevented from using chatbots to respond to their clients, though this might reduce the amount of harassment to which they are exposed \citep{10.1145/3555650}. 
Sex workers would also not be able to create consensual intimate images, which may inadvertently distort the market for images of their likenesses such that it will be dominated by non-consensual intimate images rather than images they create themselves \citep{cole2023riley}.

In a similar vein, restrictions on generating content related to illicit substances may undermine harm reduction initiatives \citep{EZELL2024568, LOVEROCK2023110878}.
Character.AI's acceptable use policy states that ``[y]ou agree not to submit any Content that...seeks to buy or sell illegal drugs'', while four other developers' policies prohibit content related to illicit substances (Anthropic, Meta, Google, OpenAI). 
These restrictions impact not only organized criminal groups seeking to scale-up mass distribution of illicit substances,
but also social services organizations that follow best practices by working to promote harm reduction rather than abstinence for populations with substance use disorders \citep{SAMHSA2023HarmReduction, marlatt1996harm}. 

These potentially beneficial uses of generating prohibited content should lead developers to weigh the costs and benefits of including and enforcing each prohibited use in their acceptable use policies. 
Some developers may choose to not enforce their policies in risky domains that could present benefits (e.g., robotics and harm reduction), making their policies less stringent in practice. 
But many developers reuse acceptable use policies from other organizations, promoting standardization while reducing the likelihood that each provision will be carefully considered \citep{mcduff2024standardization}. 
Marginalized populations such as sex workers may be harmed by disproportionate policy enforcement \citep{strohmayer2019technologies}. \\

\noindent \textbf{5.3 Lack of transparency in enforcement}

\noindent There is little publicly available information about how acceptable use policies are enforced \citep{shahi2024generative}. 
Although companies make the prohibited uses of their models clear, it is unclear how they enforce their policies in practice. 
Foundation model developers provide little or no information about how they respond to policy violations, or whether they provide justification or appeals processes when they do so \citep{bommasani2023foundation}. 
\cite{bommasani2024foundationmodeltransparencyindex} compile and release transparency reports from foundation model developers, finding that 10 of 14 disclosed some high-level details related to enforcement, though just 8 disclosed if they allow users to appeal decisions and 7 disclosed if justification is provided when enforcement occurs; notably, Google disclosed it has not taken any enforcement actions under its Generative AI Prohibited Use Policy \citep{weidinger2024starsociotechnicalapproachred}. 

This lack of transparency is different from other digital technologies; social media companies, for instance, regularly release transparency reports that provide details about how they enforce their acceptable use policies and other provisions in their terms of service \citep{kaushal2024automated}. 
Still, as \cite{douek2022content} writes, ``it’s hard to overstate both how ineffective platforms are at enforcing their rules, and how little is known about what systems they have in place to do so.''
Companies are moving quickly to deploy foundation models at the same time as they have downsized the trust and safety teams required to enforce acceptable use policies \citep{motyl2024}.

Without information about how acceptable use policies are enforced, it is not evident that they are currently being implemented or effective in limiting dangerous uses of foundation models \citep{bommasani2024foundation}. 
Some firms may publish acceptable use policies as a type of public relations statement to demonstrate they are responsible organizations, as firms incur no costs for doing so if they do not invest in enforcement \citep{Floridi2019}. 

\section{6. Discussion}
\noindent \textbf{6.1 Developers decide what constitutes acceptable use}

\noindent Acceptable use policies are written by developers without input from users or external partners. 
Developers alone have the ability to decide how their foundation models and the AI systems that integrate them are used; foundation models sit at the center of generative AI supply chain, granting developers outsized power in this ecosystem \citep{bommasani2022opportunities, cen2023ai}.
While corporate users may negotiate more permissive licenses, individual users have no means of negotiating changes to the terms of service. 
Foundation model developers with acceptable use policies include some of the world’s largest companies (e.g., Amazon, Google, Meta, Microsoft), and their choice of what constitutes unacceptable use stems in large part from the need to reduce their own legal, political, and reputational risks, not the risks to their users \citep{gillespie_custodians_2018}.

Since 2023, developers have made some effort to broaden the group of people responsible for determining the boundaries of acceptable use. 
For instance, \cite{10.1145/3630106.3658979} conducted a survey of Americans to solicit their views regarding how language models should behave, then updated the model behavior policy for an Anthropic model by using respondents' preference data during fine-tuning. 
This is part of what \cite{10.1145/3617694.3623261} call the ``participatory turn in AI design,’’ with some developers suggesting they may incorporate surveys into policy development \citep{suresh2024participation, 10.1145/3551624.3555290}.
Open-weight foundation models without use restrictions also widen the circle of who can be involved in such decisions, providing an opportunity for downstream developers to choose different acceptable use policies and adapt the model such that it is more likely to comply with the policy \citep{bommasani2023governing}. 

But these efforts to broaden participation in policy design fall short of addressing the lack of legitimacy that firms may face in deciding how an entire class of new general-purpose technologies may be used \citep{suresh2024participation, 10.1145/3613904.3642775, 10.1145/3630106.3658973, doi:10.1177/20539517241248093, madison2003reconstructing, douek_2024}.
Technology companies were not chosen to be the arbiters of what AI-generated content is acceptable by a democratic process \citep{Gautam2024, seger2023democratisingaimultiplemeanings}; rather, as \cite{Ovadya2023} writes, powerful corporations that ``unilaterally control extraordinarily powerful AI systems'' may represent a form of ``autocratic centralization.''
Several of the largest foundation model developers are currently facing antitrust lawsuits in the US which allege they broke the law to obtain their dominant market position \citep{us_v_google_2024, vestager_cardell_kanter_khan_2024}. 
The oligopoly in the cloud market limits the ability of startups and competitors to develop and distribute foundation models without the influence of incumbents, further concentrating decision-making power over what constitutes acceptable use \citep{cma_ai_2023, cma_ai_update_2024, hu_bensinger_godoy_2024, widder2023open, lehdonvirta2022cloud}.

Developers' enforcement of acceptable use policies for foundation models is likely to suffer from many of the issues  digital platforms face in enforcing their content policies \citep{gillespie_custodians_2018}. 
Social media companies are regularly accused of disparate and unequal enforcement of their policies, amplifying white supremacist, misogynist, and far-right content while enforcing their policies against Muslims, people of color, and dissidents \citep{haimson2021disproportionate, siapera2021governing, donovan2022meme}. 
Marginalized communities have fewer resources for advocacy to persuade firms that their content should be considered ``acceptable,'' meaning that centralized decision-making regarding policy enforcement often reinforces majoritarian views \citep{solaiman2024evaluatingsocialimpactgenerative}.
\\

\noindent \textbf{6.2 Gaps in use restrictions may facilitate misuse}

\noindent Developers’ acceptable use policies have substantial differences in key areas. While many developers restrict content related to politics and medical advice, more than two-thirds of developers have no such prohibitions. And while some companies’ policies prevent their models from being used by content farms or the legal services industry, some have few industry-related restrictions and others release noncommercial models with no other restricted categories of use.

The lack of consistency across developers' acceptable use policies could facilitate misuse in three ways.
First, it makes policy enforcement more difficult. 
Different policies may require different enforcement mechanisms; for example, building a filter for prompts related to glorifying violence requires different data (e.g., blocklists) than for prompts related to producing malware \citep{jhaver2018online}. 
As a result, it is more difficult for deployers to enforce the acceptable use policies for models on their platforms, creating opportunities for deliberate misuse. 
It is also unclear how to properly combine two acceptable use policies for different models \citep{villa2022evaluating}, as would be needed in the case of a model that makes use of multiple other models, as with model merging \citep{choshen2022fusingfinetunedmodelsbetter}, mixture-of-agents \citep{wang2024mixtureofagentsenhanceslargelanguage}, or other systems in which an agent interacts with other models \citep{lai2024position}.
And if the outputs of a model are used as part of the training data of another model, the latter model might include data that does not reflect its acceptable use policy if the two models’ policies differ. 

Second, the lack of consistency diminishes users’ understanding of what uses of a foundation model are acceptable. 
Many users regularly interact with multiple foundation models, such as the voice assistant on their smartphone, the summarization model in their search engine, and a standalone chatbot for brainstorming or coding. 
Each of these models may have a different acceptable use policy, meaning the average user may struggle to internalize which uses are disallowed \citep{doi:10.1080/1369118X.2018.1486870}.
This is likely to produce accidental misuse. 

Third, models are not safety-tuned for less common restricted uses. 
Without strong norms in the developer community about which uses are unacceptable, developers are less likely to invest resources in making their models refuse to generate related content \citep{reuel2024openproblemstechnicalai}.
As a result, there is a lack of data that developers can use to build filters for less common prohibited use categories, such as self-harm.

Every acceptable use policy need not be the same, but the lack of standardization is creating negative externalities in the ecosystem.
At minimum, developers could work to build consensus around what constitutes acceptable use and aim to make their policies interoperable where appropriate. \\ 

\noindent \textbf{6.3 AUPs help shape the foundation model market}

\noindent Acceptable use policies alter the foundation model market by affecting which organizations can use a model and for what purpose. For example, developers use these policies to prevent companies from making use of their services, stealing their intellectual property, or building a competing model. 
Many companies ban firms and other users from using their models to train other machine learning models, restricting the supply of datasets of model outputs and concentrating the market for models that are trained on their model’s outputs \citep{zhao2024wildchat1mchatgptinteraction}. 
On the other hand, in July 2024 Meta updated a license to allow users to use outputs from Llama 3.1 to ``to create, train, fine tune, or otherwise improve an AI model,'' perhaps in an effort to gain market share \citep{lambert_2024}.

Acceptable use policies also help determine what industries can make use of developers’ models. 
Policies that prohibit the use of models for weapons production may block the arms industry from making use of those foundation models, as with surveillance tech companies and political advocacy groups.
These policies also determine the types of uses of models that are permitted (e.g., no automated decision systems, no automated posting of AI-generated outputs). 
Even industries that are allowed to make use of models may not be able to do so for common applications.

\section{7. Areas for Future Work}
\noindent  \textbf{7.1 Collecting data on AUP enforcement} \\
\noindent The way in which developers and deployers enforce acceptable use policies for foundation models remains unclear.
Collecting data related to enforcement is a key area for future work, as there is little indication that companies will share quantitative data regarding enforcement \citep{bommasani2024foundation}.
This data might be collected by asking users to donate their data (e.g., chat logs), surveying users about their experiences, or working with companies to gain access.
One key question is how enforcement differs depending on the system the foundation model is embedded within; for instance, some companies might enforce their acceptable use policy less strictly for language models distributed via API as opposed to via a chat interface, as there are more users of chatbots. \\

\noindent \textbf{7.2. Content moderation on generative AI platforms}
\noindent Content moderation has been studied much more thoroughly on social media platforms than on AI platforms despite the fact that researchers have access to foundation models but lack access to underlying recommendation systems \citep{10.1145/3630106.3658932}. 
Some foundation model developers have adopted content moderation practices quickly, hiring trust and safety teams and adopting acceptable use policies to curb undesirable content. 
The same scrutiny that is applied to content moderation on social media should be applied to developers' enforcement of acceptable use policies, including the data labor employed as part of this work \citep{gray2019ghost, roberts_behind_2019}. 
Evaluations of foundation models can also be seen as a form of content moderation, as they are used to assess whether a model will produce violative content and inform interventions to reduce this behavior \citep{nangia2020crowspairschallengedatasetmeasuring, zhao2018genderbiascoreferenceresolution}. \\

\noindent \textbf{7.3. Regulating acceptable use of foundation models}

\noindent Governments have attempted to spur self-regulation in the foundation model ecosystem through voluntary codes of conduct \citep{wh2023voluntary}. 
The extent to which governments can go further by forcing developers to block foundation models from generating certain types of content will likely be decided in the courts, much as the ability of social media companies to carry out content moderation has been challenged in the US \citep{netchoice_v_paxton_2024}. 
Still, China has been somewhat successful in forcing developers to censor the outputs of their foundation models, and many other governments may follow suit \citep{zeng2024airbench2024safetybenchmark}. 
Whether it is feasible for developers to follow such mandates remains an open question, given the ease with which downstream developers can remove safety mitigations.

\section{8. Conclusion}
This paper finds that there is significant heterogeneity across acceptable use policies, that they help shape the market for foundation models, and that developers adopt them in order to reduce legal and reputational risks.
There is little transparency about acceptable use policies, but this paper takes a first step towards shining a light on why they matter.

\newpage 
\section{Acknowledgments}
I thank Ahmed Ahmed, Rishi Bommasani, Peter Cihon, Evelyn Douek, Carlos Muñoz Ferrandis, Peter Henderson, Daniel Ho, Aspen Hopkins, Sayash Kapoor, Percy Liang, Shayne Longpre, Emma Lurie, Daniel McDuff, Aviya Skowron, Dave Willner, Betty Xiong, and Yi Zeng for feedback and discussions on this work. All errors and omissions are my own.

\bibliographystyle{plainnat}
\bibliography{ref}
\end{document}